\documentclass[showpacs,aps,prl,twocolumn,floatfix]{revtex4}
\usepackage{CJK}
\usepackage{graphicx}
\usepackage{epsfig}
\usepackage{wasysym}
\usepackage{color}
\newcommand{\sumnear}{\mathop{\sum}_{\langle i j \rangle}}
\newcommand{\idx}[1]{{_{\mathrm{#1}}}}

\newcommand{\iket}[2]{| #1 \rangle \idx{#2} }

\newcommand{\ifbra}[2]{\langle #1 |\idx{#2}}

\newcommand{\euler}[1]{\mathrm{e}^{#1}}
\newcommand{\imagu}[0]{i}
\newcommand{\QD}[0]{D}

\begin{document}
\title{The phase diagram of the antiferromagnetic XXZ model on the triangular lattice}
\author{Daniel Sellmann}
\affiliation{Physics Dept.~and Res.~Center OPTIMAS, Univ.~of
Kaiserslautern, 67663 Kaiserslautern, Germany}
\author{Xue-Feng Zhang }
\thanks{Corresponding author: zxf@physik.uni-kl.de}
\affiliation{Physics Dept.~and Res.~Center
OPTIMAS, Univ.~of Kaiserslautern, 67663 Kaiserslautern, Germany}
\author{Sebastian Eggert}
\affiliation{Physics Dept.~and Res.~Center OPTIMAS, Univ.~of
Kaiserslautern, 67663 Kaiserslautern, Germany}
\date{\today}

\begin{abstract}
We determine the quantum phase diagram of the antiferromagnetic spin-1/2
XXZ model on the triangular
lattice as a function of magnetic field and anisotropic coupling $J_z$.
Using the
density matrix renormalization group (DMRG) algorithm in two dimensions we establish
the locations of the phase boundaries between a plateau phase
with 1/3 N\'eel order and two distinct coplanar phases.  The two coplanar phases are
characterized by a simultaneous breaking of both
translational and U(1) symmetries, which is reminiscent of supersolidity.
A translationally invariant umbrella phase
is entered via a first order phase transition
at relatively small values of $J_z$ compared to the corresponding
case of ferromagnetic
hopping and the classical model.
The phase transition lines meet at
two tricritical points
on the tip of the lobe of the plateau state, so that the two coplanar states are
completely disconnected.
Interestingly, the phase transition between the plateau state and the upper
coplanar state changes from second order to first order for large values of $J_z \agt2.5J$.
\end{abstract}



\maketitle

\begin{figure}[t]
\begin{center}
\includegraphics[width=\columnwidth]{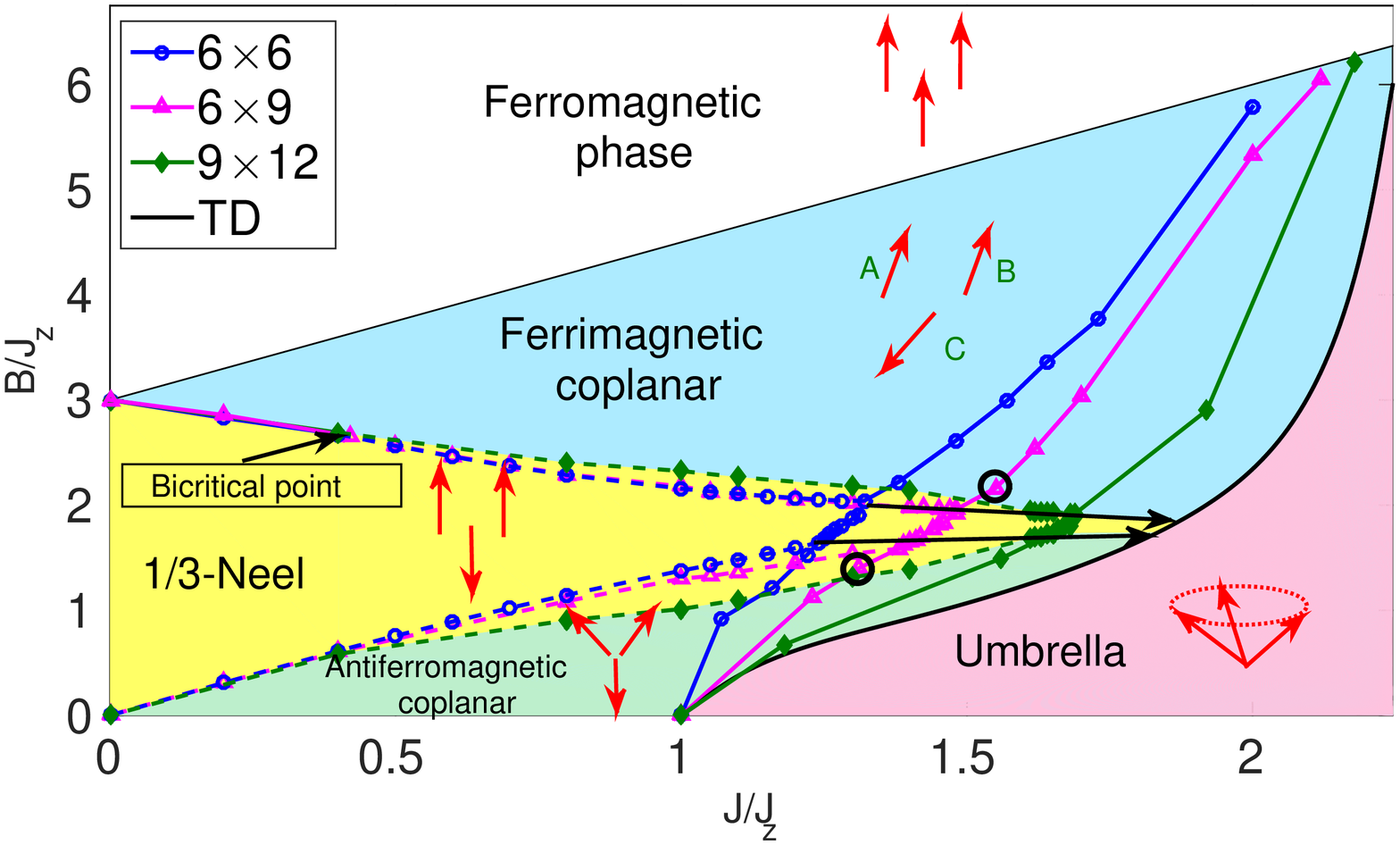}%
\caption{(Color online) {The phase diagram of the XXZ model on the triangular lattice with two dimensional DMRG and exact diagonalization
\cite{highfield} for different sizes. The solid (dashed) line represents first (second) order phase transitions, respectively. 
Arrows indicate the classical spin
configurations in the different phases. The black circles indicate the regions analyzed in
Fig.~\ref{ss}. Linear finite size scaling with $1/N$ of
the interpolated data predicts the black solid line as the phase boundary 
in the thermodynamic limit (TD).
The two black arrows show the finite size scaling 
of the tri-critical points (see the appendix for details.)}
\label{dmrg}}\end{center}
\end{figure}

Competing interactions between quantum spins can prevent conventional magnetic order
at low temperatures. In the search of interesting and exotic quantum phases
frustrated systems are therefore at the center of theoretical
and experimental research in different areas of physics \cite{balents10,anderson,sorella99,ed,ccm,tri11,tri10,tri12,tri13,
watarai01,tocchio13,balents13,white11,tocchio14,miyashita10,seabra11,kawamura85,classical,xxz,yamamoto2,
starykh14,tri,moessner08,imp,tri_1st,tri_sc,tri_af,metav14,yoshikawa04,zhang13,tri_afn,jiang08}.
One of the most straight-forward frustrated system
is the spin-1/2 antiferromagnet (AF) on the triangular lattice, which was also the first model
to be discussed as a potential candidate for spin-liquid behavior without conventional
order by  Anderson \cite{anderson}. It is now known that the
isotropic Heisenberg model on the triangular lattice is not a spin liquid and
does show order at zero temperature \cite{sorella99}.  Nonetheless,
the phase diagram as a function of magnetic field is still actively discussed with recent
theoretical calculations \cite{ed,ccm} as well as
experimental results \cite{tri11,tri10,tri12,tri13} on Ba$_3$CoSb$_2$O$_9$, which appears
to be very well described by a triangular AF.
Interesting phases have also been found for anisotropic triangular lattices \cite{tocchio13,balents13,white11} and
for the triangular extended Hubbard model \cite{tocchio14}.
{ Hard-core bosons with nearest neighbor interaction 
on a triangular lattice correspond to the xxz model with ferromagnetic 
exchange in the xy-plane, which has been studied extensively \cite{tri,moessner08,imp,
tri_1st,tri_sc,tri_af}.  
In this case a so-called supersolid phase near half-filling has been established for large
interactions \cite{tri}, 
which is characterized by
{\it two} order parameters, namely a superfluid density and a $\sqrt{3}\times\sqrt{3}$
charge density order.  
Impurity effects show that the two order parameters are competing \cite{imp} and 
the transition to the superfluid state is first order \cite{tri_1st,tri_sc}.  

However, surprisingly little attention has been paid to the role of
an {\it antiferromagnetic} anisotropic exchange interaction away from 
half-filling \cite{classical,xxz,yamamoto2,starykh14},
even though the XXZ model on the triangular lattice
\begin{eqnarray}
H=J\sumnear(\hat{S}_{i}^x\hat{S}_{j}^x+\hat{S}_{i}^{y}\hat{S}_{j}^y)+J_z \sumnear
\hat{S}_{i}^z \hat{S}_{j}^z-B\mathop{\sum}_i\hat{S}_i^z,
\label{spinmodel}
\end{eqnarray}
is arguable one of the most fundamental
examples of frustrated antiferromagnetism.  Only very recently 
the first complete phase diagram as a function of $B$ and $J/J_z$ was published by 
 Yamamoto {\it et~al.}~using the cluster mean-field theory (CMF) \cite{xxz}. 
In this case, {\it three} phases with broken sublattice symmetry and simultaneously broken U(1) symmetry 
were found, which are stable to surprisingly large $J/J_z$ compared to the corresponding
ferromagnetic model.  One of those phases -- the so-called $\pi$-coplanar 
phase -- was not expected to exist at all from simple mean field considerations \cite{xxz} 
and therefore
deserves special attention.

We now present quantum simulations of this model using the 
density matrix renormalization group (DMRG)
\cite{white,schollwoeck,xiang} algorithm in  two dimensions. 
The resulting phase diagram as a function of $B$ and $J/J_z$
is summarized in Fig.~\ref{dmrg}, which first of all confirms several aspects of the
previous study in Ref.~[\onlinecite{xxz}]:
For large values of $J/J_z$, we find an {\it umbrella state} with
spontaneously broken U(1) symmetry, but no broken sublattice symmetry.
With increasing $J_z$ and at small magnetic
fields a first order transition occurs to a {\it antiferromagnetic coplanar phase}
 where the spins
on one sublattice align against the field, while the other two
sublattices form a honeycomb structure with spins still partially
pointing in the xy-plane, so that all spins lie in a plane.  
At large fields a {\it ferrimagnetic coplanar 
phase} is found with parallel canted spins on two sublattices and
one sublattice pointing in a different direction. 
A {\it 1/3 N\'eel phase} with fixed magnetization separates the two co-planar phases.
The phase transition to the saturated phase occurs exactly at $B=3(J_z+J/2)$ as for the
classical triangular
antiferromagnet \cite{kawamura85,xxz,classical,seabra11,miyashita10}. 

Our results also show several differences 
to the previous study \cite{xxz}:
1.) The so-called $\pi$-coplanar phase is missing.  
As shown below this phase exists only for small system sizes or clusters. 
2.) Two tri-critical points, which separate the 1/3-Neel phase from the umbrella phase
 are pushed to much larger values of $J/J_z$ and become very close in the thermodynamic 
limit. 
3.) The second order phase
transition between the 1/3 N\'eel phase and the ferrimagnetic coplanar
phase curiously turns first order for strong interactions $J/J_z
\alt 0.4$ at a special bi-critical point, which has since been confirmed \cite{yamamoto2}.
Similar bi-critical points where a phase transition changes from 1st to 2nd order were recently under discussion in binary Bose mixtures \cite{yamamoto3}.}

We now discuss the detailed numerical DMRG data at selected points
in the phase diagram.
Frustrated systems are known to be sensitive to 
boundary induced behavior \cite{zhang13}, so that periodic boundary conditions (PBC)
{ turned out to be necessary} in both directions \cite{tri_afn,jiang08}.  
Accordingly, the initial truncation error 
may be as high as $10^{-5}$ which is normal for 2D DMRG with PBC \cite{tri_afn,jiang08}. 
In fact, DMRG ``sweeping'' improves the data significantly (up to 16\%), so that the initial 
truncation error becomes irrelevant as a measure (which is in fact not very sensitive to $m$).
The final energy values after sweeping go to a unique value for large $m \agt 1000$, so that convergence can be ensured.
Note, that the DMRG operates in the canonical ensemble, i.e.~the data is given as a function of magnetization per site $M$ and the corresponding fields can be obtained as the derivative of the ground state energy $E(M)$ with respect to
$M$,~i.e. $B(M)=E(M+1/N)-E(M)$ \cite{sq,anyon,gap}.
The upper tricritical point can be
found by the condition
$B(1/3)=B(1/3-1/N)$.  
{There is no particle-hole symmetry so the kept states we can afford is $m=3000$ at most. Technical details about convergence and finite size scaling can be found in the 
appendix.}

\begin{figure}[t]
\begin{center}
\includegraphics[width=\columnwidth]{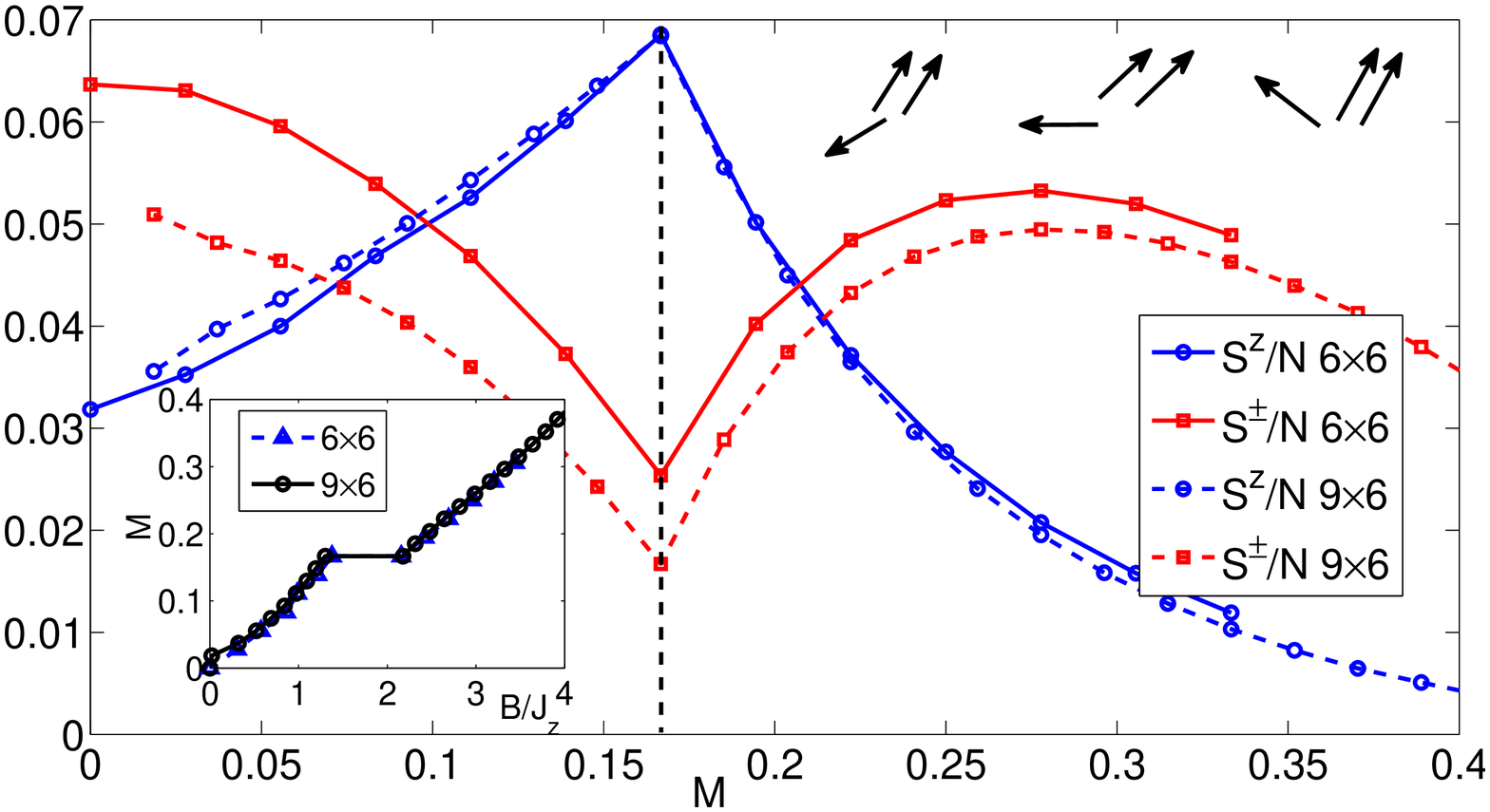}%
\caption{The structure factors $S^z(\textbf{Q})/N$ and
$S^{\pm}(\textbf{Q})/N$ at $J_z=J$ as a function of magnetization for
different sizes. The dotted line at $M=1/6$ indicates the location of the 1/3-N\'eel phase.
Inset: The magnetization as a function of magnetic
field.\label{j1}}\end{center}
\end{figure}

The Heisenberg system $J=J_z$ in a field has previously been
considered using exact diagonalization
\cite{ed,miyashita86,pierre1994,honecker1999,honecker2004}, spin
waves \cite{zheng-06,chubukov91} and coupled cluster methods (CCM)
\cite{ccm}. It is well known that the uniform magnetization has a
plateau at $M=1/6$ which is characteristic of the 1/3 N\'eel phase
as shown in the inset of Fig.~\ref{j1}. 

The structure factors in the z-direction $S^z(\textbf{Q})=\langle
|\mathop{\sum}_{k=1}^N S_k^z e^{\emph{\textbf{i}} \textbf{Q}
\cdot\textbf{r}_k}|^2\rangle/N$ and in the xy-direction
$S^{\pm}(\textbf{Q})=\langle |\mathop{\sum}_{k=1}^N S_k^+
e^{\emph{\textbf{i}} \textbf{Q} \cdot\textbf{r}_k}|^2\rangle/N$ at
$\textbf{Q}=(4 \pi/3,0)$ are useful order parameters to measure
the diagonal and the off-diagonal order, respectively.  If $S^z/N$
is finite the system has a broken sublattice symmetry (charge
order), while a finite $S^\pm/N$ indicates a broken U(1)
rotational symmetry (superfluidity).  As shown in
Fig.~\ref{j1} both order
parameters are finite in the ferrimagnetic and antiferromagnetic coplanar 
phases. At zero magnetization $S^\pm/N$ is larger than $S^z/N$,
but then decreases with $M$ and scales to zero with $1/N$ at
$M=1/6$, which is exactly the point where $S^z$ becomes largest.
In the experiments on Ba$_3$CoSb$_2$O$_9$ an additional cusp in
the susceptibility was observed at higher magnetization $M\approx
1/3$ \cite{tri11}, which could indicate another phase transition.
However, our data does not show any other phase for $M>1/6$ and
$J=J_z$. Nonetheless, the off-diagonal structure factor $S^\pm$
does show a broad maximum around $M\approx 1/3$, which is due to
the fact that the spins on one of the sublattices are able to align
along the xy-plane at approximately this magnetization as shown in
Fig.~\ref{j1}.  Spins that are aligned within the xy-plane have in
turn the largest susceptibility in the z-direction, so this could
in part explain the observed maximum in Ref.~[\onlinecite{tri11}].

\begin{figure}[t]
\begin{center}
\includegraphics[width=\columnwidth]{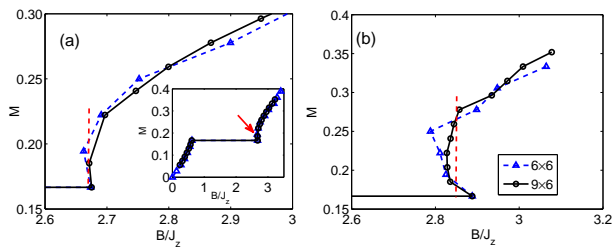}%
\caption{(Color online)
The magnetization as a function of field at (a) $J_z=2.5J$ (inset: larger range)
and (b)  $J_z=5J$ for different sizes. The dotted vertical line (red) indicates the Maxwell 
construction.
\label{j4}}\end{center}
\end{figure}

We now turn to larger values of $J_z =2.5J$, where the magnetization plateau is
larger {than for $J_z =J$} as shown in the inset of  Fig.~\ref{j4}.  The behavior of the order parameters
$S^\pm$ and $S^z$ is qualitatively similar to the isotropic case as
a function of magnetization.  However, for the phase transition between the 
1/3 N\'eel phase and the 
{\it ferrimagnetic} coplanar
there is a subtle, but important difference in the
magnetization curve at strong interactions.
 As shown in Fig.~\ref{j4},
near the upper
phase boundary the calculated field decreases
with increasing magnetization (which is fixed for each simulation).
This behavior indicates an unstable state
and in the thermodynamic limit leads to 
phase separation, which is an obvious indication of a first order phase transition.
In a finite system the energy of the phase boundary can prevent phase separation
and the unstable state can be found by numerical simulations at a given magnetization, 
which is the case
here and in related systems \cite{poilblanc09,sq,yamamoto2}.
The corresponding first order jump in magnetization must then be determined
by a Maxwell construction as indicated in Fig.~\ref{j4}. 
This jump vanishes somewhere between
$J_z =2.5J$ and  $J_z=2J$, so that we predict a bicritical point where the
second order phase transition turns first order in the strong coupling
limit as shown in Fig.~\ref{dmrg}.
This surprising behavior can in part be explained
from the fact that the end of the $M=1/6$ plateau approaches
the saturation field, so that there is only a small field region where
the magnetization changes from $M=1/2$ down to $M=1/6$.  However, the coplanar spin
state has only a limited susceptibility close to saturation, so that a jump in magnetization
may be the only way to resolve this contradiction.  In other words, starting from the
1/3 N\'eel state the configuration must make a finite jump to reach the coplanar state
if the upper critical field is too large, since the ferrimagnetic coplanar state
is already canted significantly towards the field in this case.
In any case, the quantum mechanical mechanism for this behavior is an interesting
aspect for future studies.
The second order phase transition between the
antiferromagnetic coplanar phase and the plateau phase is well
understood from a
strong coupling expansion \cite{tri_sc} in terms of holes which start to occupy
the honeycomb sublattice at a critical value of $B\approx 3J/2 +5 J^2/8J_z - 71J^3/32J_z^2$, which is consistent with our numerical data.

{We should emphasize that order parameters do not have to be
used in order to determine the phase transitions
from the 1/3-N\'eel phase to the coplanar states, since the magnetization plateau can 
be determined directly from the energies $E(M)$.
In order to study the phase boundaries to the umbrella  phase, on the other hand,
 order parameters 
are essential, but this becomes numerically
costly for large system sizes.
As additional tools, we therefore 
want to explore here if different measures of entanglement and quantum discord are
useful in 2D DMRG, which have been
proposed and used for studying quantum phase transitions in recent related systems \cite{ee1,ee2,ee3,Wootters,dillen08,QD}.
To define suitable quantum information measures}
it is useful to consider
the reduced density matrix $\rho_{ij}$
of two neighboring spins. 
 The trace over spin $j$ gives the
reduced density matrix of a single spin  $\rho_i = \mathrm{Tr}_j \rho_{ij}$.
The von-Neumann entropy of a general density matrix $S_A=-\mathrm{Tr}\rho_A \log \rho_A$
can be used to define the entanglement entropy $S_i$.
The concurrence \cite{Wootters,dillen08}
\begin{equation}
C_{ij}= 2 {\rm max}(0, \sqrt{\lambda_1}-\sqrt{\lambda_2}-\sqrt{\lambda_3}-\sqrt{\lambda_4}),
\label{cc}
\end{equation}
is given in terms of the eigenvalues $\lambda_i$ of the matrix $\rho_{ij} \tilde{\rho}_{ij}$,
where $\tilde{\rho}_{ij}$ characterizes the spin-flipped state.
The quantum discord \cite{QD,dillen08} has
been proposed as a good indicator for quantum phase transitions
\begin{equation}
\label{eqQDDef}
\QD_{ij}=\mathrm{min}_{\lbrace \Pi^j_\nu
\rbrace}\left(S_{i}-S_{ij}+S_{i|j} \right),
\label{QD}
\end{equation}
which is calculated in terms of the conditional quantum entropy
\begin{equation}
S_{i|j}
=\sum_{\nu=1}^2 p_\nu
S(\rho_{i|\Pi^j_\nu}),
\end{equation}
where
$\rho_{i|\Pi^j_\nu}=\Pi^j_\nu \rho_{ij} \Pi^j_\nu$
and $p_\nu=\mathrm{Tr} \Pi^j_\nu \rho_{ij} $.
{The projectors $\Pi_\nu=\iket{\psi}{\nu}\ifbra{\psi}{\nu}$
can be defined in terms of a general parametrization}
\begin{eqnarray}
\iket{\psi}{1} & = &\cos{\theta} \iket{\downarrow}{j} + \euler{-\imagu\phi} \sin{\theta}
\iket{\uparrow}{j} \nonumber\\
\iket{\psi}{2} & = &\euler{+\imagu\phi}\sin{\theta} \iket{\downarrow}{j} -  \cos{\theta}
\iket{\uparrow}{j} . \label{theta}
\end{eqnarray}
The minimization over the projectors in (\ref{eqQDDef}) then corresponds to a
minimization over angles $\theta$ and $\phi$ in the wavefunctions.
\begin{figure}[t]
\begin{center}
\includegraphics[width=\columnwidth]{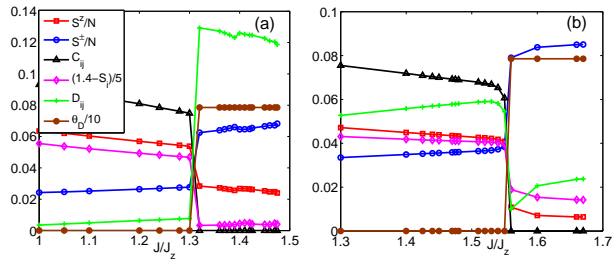}%
\caption{The structure factor $S^z(\textbf{Q})/N$,
$S^{\pm}(\textbf{Q})/N$, concurrence, entanglement entropy, quantum discord and
the variational angle $\theta$ in the $6\times9$ lattice in the regions
shown by red circles in Fig.~\ref{dmrg} indicate a
phase transition to the umbrella state from (a) the
antiferromagnetic ($B=1.398 J_z$) and (b) the ferrimagnetic coplanar state ($B=2.161J_z$).
\label{ss}}\end{center}
\end{figure}

\begin{figure}[t]
\begin{center}
\includegraphics[width=\columnwidth]{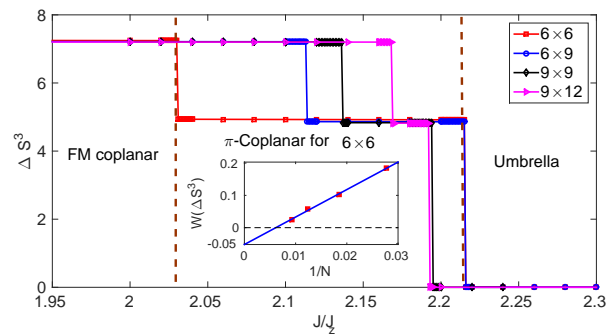}%
\caption{The order parameter $\Delta S^3$ in Eq.~(\ref{Delta})
as a function of $J/J_z$ close to saturation $M=1/2-3/N$. Inset: finite size scaling of the $\pi$-coplanar plateau width.
\label{pi}}\end{center}
\end{figure}

In Fig.~\ref{ss} we show the two
order parameters, the concurrence, the entanglement entropy, and
the quantum discord
at two selected points in the phase diagram, which
are indicated by {black} circles in Fig.~\ref{dmrg}.
All measures give the same locations of the phase transition
(in this case $B=1.398 J_z$, $J=1.31J_z$
and $B=2.161J_z$, $J=1.55J_z$, respectively). 
The quantum information measures based on  
$\rho_{ij}$
are computationally less demanding than the structure factors since they 
can be determined
from the correlation functions of only two neighboring spins \cite{dillen08}.
They are also universal, since no particular order needs to be assumed.
In particular, the  quantum 
discord $\QD_{ij}$ \cite{QD,dillen08} turns out to be very reliable
in detecting the phase transitions and interestingly the 
corresponding variational angle $\theta$ in Eq.~(\ref{theta})
 takes on different values on the two sides of the phase transition.
It is so far unclear if this jump in a variational parameter is a generic feature, but
it may be useful in future studies as well.
We find that the phase transition between the ordered states (N\'eel and coplanar)
to the umbrella phase
is always first order, except at the isotropic point $B=0$, where it is known to
be second order \cite{tri_afn}.
At two tricritical points
the second order phase transitions
between N\'eel and coplanar phases meet the first order transition. 
{The phase transition to the umbrella phase can be accurately determined for 
system sizes of up to $9\times 12$, so that a systematic finite size scaling becomes
feasible. 
The (interpolated) first order 
phase transitions to the umbrella phase can be linearly extrapolated
in $1/N$ which gives an estimate in the thermodynamic limit (TD) shown
in Fig.~\ref{dmrg}.  
Extrapolating with a different power $1/\sqrt{N}$ also gives a reasonable fit and  pushes
the phase transition line out even further by up to $0.3J/J_z$, which would make an even
larger quantitative difference to the coupled cluster study \cite{xxz}.  The 
two tri-critical points approach each other with finite size scaling and we cannot rule 
out that they merge to one single multicritical point in the TD.
While finite size scaling works reasonably well
for the first order phase transition, the same is not true for the second
order phase transition lines, which show a much more irregular behavior with 
system size, that we cannot explain.

Finally, we have made a focused search using exact diagonalization \cite{highfield}
for the new ``$\pi$-coplanar'' phase, which was postulated in Ref.~[\onlinecite{xxz}].} 
 We found a suitable order parameter to be 
\begin{equation}
\Delta S^3=N^3 \langle (M_A- M
)(M_B- M )(M_C- M
)\rangle, \label{Delta}
\end{equation}
where $M_{A,B,C}$ is
the magnetization on each sublattice $A,B,C$.
The parameter $\Delta S^3$ shows { \it three different} values
in the
ferrimagnetic coplanar, the $\pi$-coplanar and the umbrella phase,  respectively as shown in Fig.~\ref{pi}.  {For small system sizes
of $9\times 12$ or less a $\pi$-coplanar phase can be identified, but it shrinks fast with increasing system size.  With finite size scaling shown in the inset of Fig.\ref{pi}, the $\pi$-coplanar phase disappears for $N \agt 200$. Therefore, we predict that there is no such phase in the thermodynamic limit.}

In conclusion, we have analyzed the spin-1/2 XXZ model on the
triangular lattice using a two dimensional DMRG method with periodic boundary conditions.
The phase diagram shows two coplanar phases with
different symmetries of the superfluid
condensate, which is separated by an ordered plateau 1/3 N\'eel phase,
with fixed magnetization $M=1/6$.
The transition to the umbrella state is always first order for finite fields
and the critical line $B_c(J)$ in Fig.~\ref{dmrg} is monotonically increasing, so that
a larger field always leads to an extended ordered state.
The transition between the coplanar and the 1/3 N\'eel phase is generically second
order but curiously the upper phase transition line turns first order
for $J_z \agt 2.5 J$, which is yet not fully understood.

\begin{acknowledgments}
We are thankful for useful discussions with Axel Pelster about his
meanfield calculation of the extended Bose-Hubbard model, Shijie Hu about numerical 
suggestions, Tao Shi and
Raoul Dillenschneider
about the spin wave calculations and comments from Alexandros Metavitsiadis, Denis Morath
and Dominik Stra\ss el. This work was supported by the ``Allianz
f\"{u}r Hochleistungsrechnen Rheinland-Pfalz" and by the DFG via
the SFB/Transregio 49.
\end{acknowledgments}

\appendix

\section{Appendix: Finite size scaling of the phase diagram}
\begin{figure}[b]
\begin{center}
\includegraphics[width=.95\columnwidth]{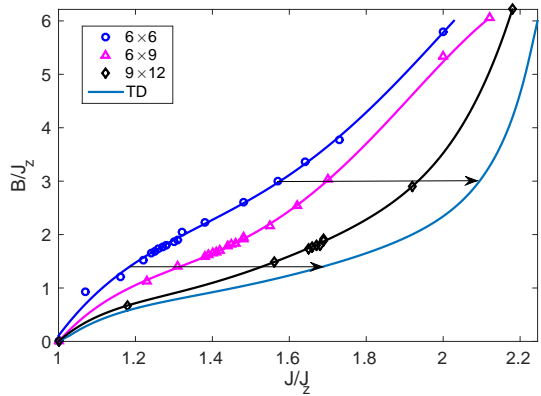}%
\caption{(Color online) The phase boundaries of the umbrella phase for different sizes. 
The black arrows indicate the two examples, for which the behavior with
$1/N$ is shown in Fig.~\ref{luffs2} (top).
\label{spine}}\end{center}
\end{figure}
\begin{figure}[t]
\begin{center}
\includegraphics[width=.83\columnwidth]{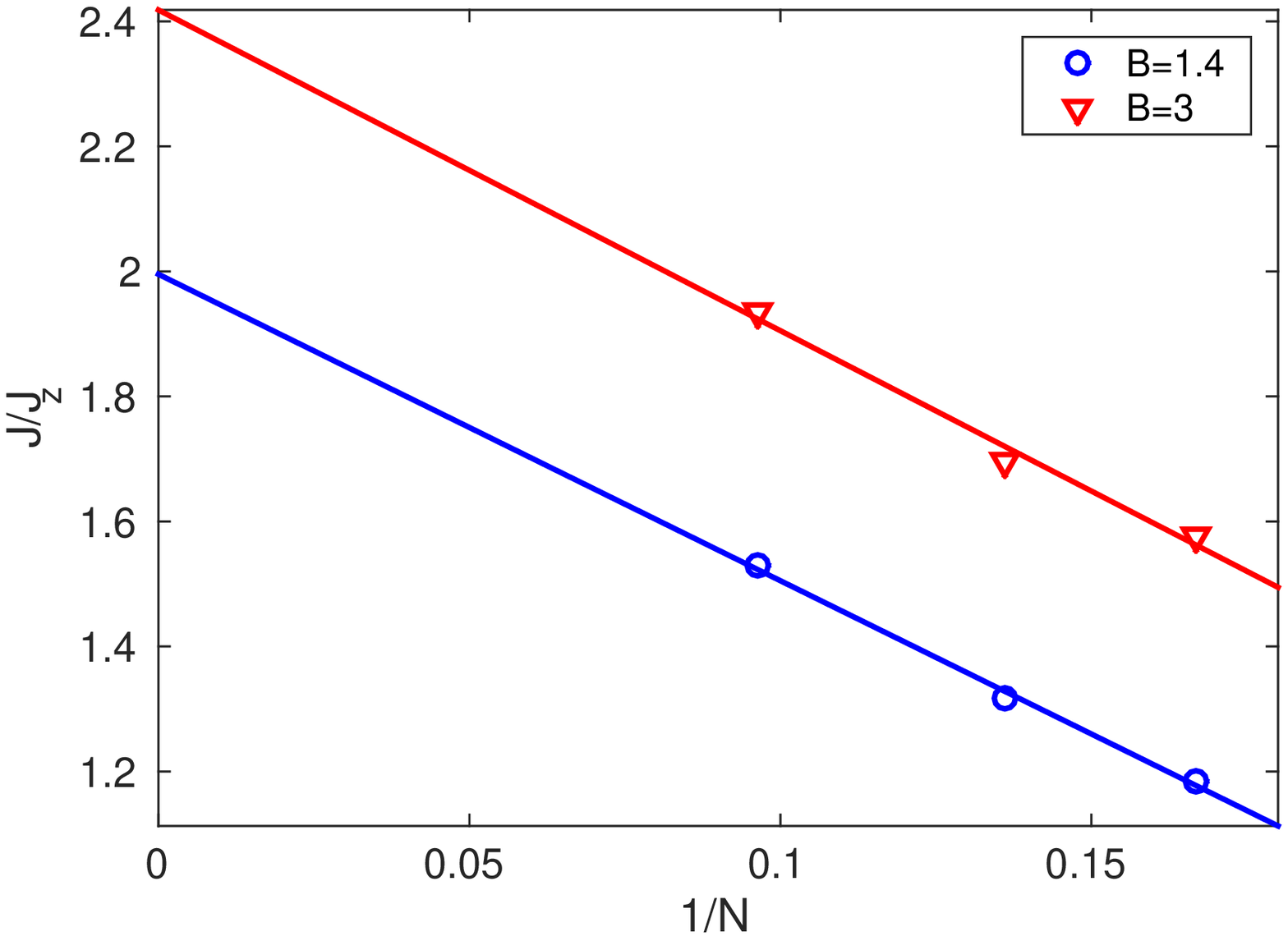}\\
\includegraphics[width=.83\columnwidth]{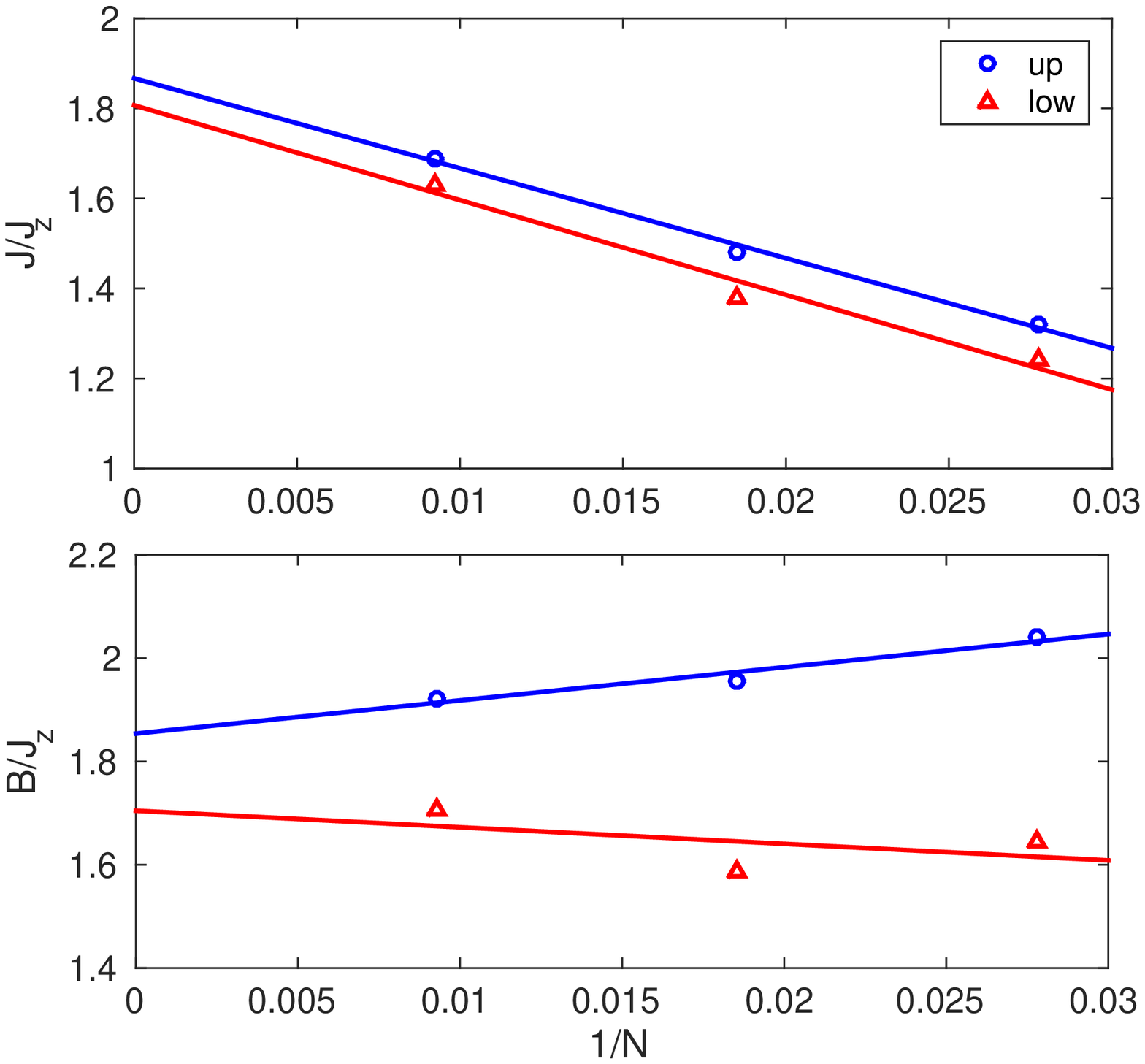}%
\caption{(Color online) Top: The finite size scaling of the first order phase transition to 
the umbrella phase at $B/Jz=1.4$ and $B/Jz=3$.
Lower two plots: The finite size scaling of the locations $J$ (middle plot) and $B$ (bottom plot) 
of the upper and lower tri-critical point.
\label{luffs2}}\end{center}
\end{figure}

The first order phase transition to the umbrella phase for a given system size can be 
determined rather accurately as shown in Fig.~4 in the main manuscript.
In those simulations the magnetization is fixed and the corresponding
magnetic field is determined by the derivative of the ground state energy
$B(M) = E(M+1/N)-E(M)$ at the transition point.  This yields the phase transition lines 
for system sizes $6\times 6$, $6\times 9$, and $9\times 12$ shown in Fig.~\ref{spine} below.
Since the data is well behaved, it is possible to determine the corresponding
continuous curves $B(J)$ for all values of $B$ by spline interpolation.
For each field it is then possible to use a linear fit in reciprocal system size $1/N$
as shown in Fig.~\ref{luffs2} (top), which determines the estimate in the thermodynamic limit
in Fig.~\ref{spine}.  A reliable error estimate is difficult in this case, since
the finite size data is only available for 
three data points and a square root behavior $1/\sqrt{N}$ also yielded reasonable fits,
which would push out the estimate of the phase transition line to higher values of $J$
by up to  $J/J_z \sim 0.3$.
Therefore, the phase transition line shown should be 
taken as a lower estimate.

For each finite size we find two tri-critical points where one coplanar, the 1/3 N\'eel,
and the umbrella phase meet.  The location of the tri-critical points change 
in both $B$ and $J$ with finite size, but a reasonable estimate of the corresponding values
thermodynamic limit can be made as shown in Fig.~\ref{luffs2} (lower two plots).  Since the two points
come quite close with finite size scaling it cannot be ruled out from our data 
that they merge into
one multi-critical point in the thermodynamic limit.

The situation is even more complicated for the {\it second order} phase transition lines
from the 1/3 N\'eel phase to the coplanar phase.  In this case we could not find any 
systematic finite size scaling, since the phase transition lines for 
$6\times 6$ and $6\times 9$ 
are very close but there is a larger change when going to $9\times 12$.   We cannot explain this
behavior, but it is maybe not surprising, that 
it is more difficult to pinpoint the transition lines for 
second order phase transitions than for first order transition lines 
in the thermodynamic limit.

\begin{figure}[t]
\begin{center}
\includegraphics[width=.8\columnwidth]{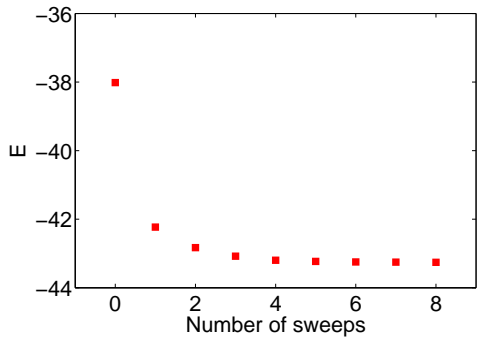}%
\caption{The energy estimate as a function of the number of sweeps for 
$J/J_z=1.3$, $M=1/6+2/N$ and $N=9\times12$ with $m=1600$ kept states.
\label{sweep}}\end{center}
\end{figure}
\section{Appendix: Convergence of the DMRG data in 2D}
In order to study two-dimensional systems with the DMRG algorithm, the sites need to be 
ordered along a one-dimensional chain with effectively long range interactions.
Therefore, neighboring sites may be rather far apart in the DMRG algorithm, so that more
effort is required to capture the quantum correlations.  This problem is even more 
severe with periodic boundary conditions in both direction, which were necessary for
the xxz model on the triangular lattice. 
As a result we find that the variational state of the system is rather poorly described
after the initial DMRG  buildup.  Typically, the truncation error is not very small 
($\sim 10^{-5}$)
and is not a reliable measure of the quality of the simulations
(in fact it does not depend much on the number of states kept).
The finite size algorithm quickly improves this 
state by ``sweeping''. While the correction in energy maybe as large as 20\% 
in the first sweep,  the changes become several orders of magnitude smaller after 
just a few iterations 
as shown in Fig.~\ref{sweep}. 

\begin{figure}[t]
\begin{center}
\includegraphics[width=.8\columnwidth]{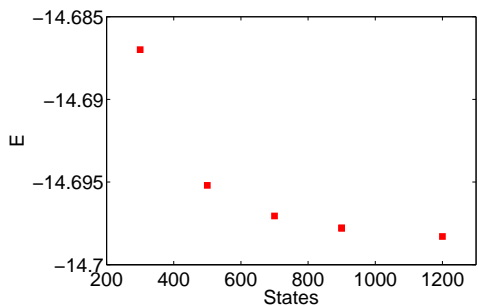}
\caption{The energy estimate as a function of the 
number of kept states $m$ for
$J/J_z=1$, $M=1/6$ and $N=6\times6$.
\label{L36e}}\end{center}
\end{figure}
However, 
the convergence with the number of sweeps is also not a guarantee 
that the system is approaching the correct ground state, since 
metastable states are possible.  It is therefore essential to vary the number 
of states kept.  It is also possible to change the number of kept states during the
sweeping procedure or change the initial buidup geometry.  Typically, we find that 
if the data has a smooth behavior with the number of states kept it also 
produces sensible and accurate 
data which fits well into the phase diagram (e.g.~relative to neighboring 
points in parameter space).
Hence each data point in the phase diagram has been carefully checked for consistency.
In principle it would also be possible to try an extrapolation fit with 
the number of states, but in practice this only produces tiny corrections
to the phase diagram but may in turn produce artifacts.

\begin{figure}[t]
\begin{center}
\includegraphics[width=.8\columnwidth]{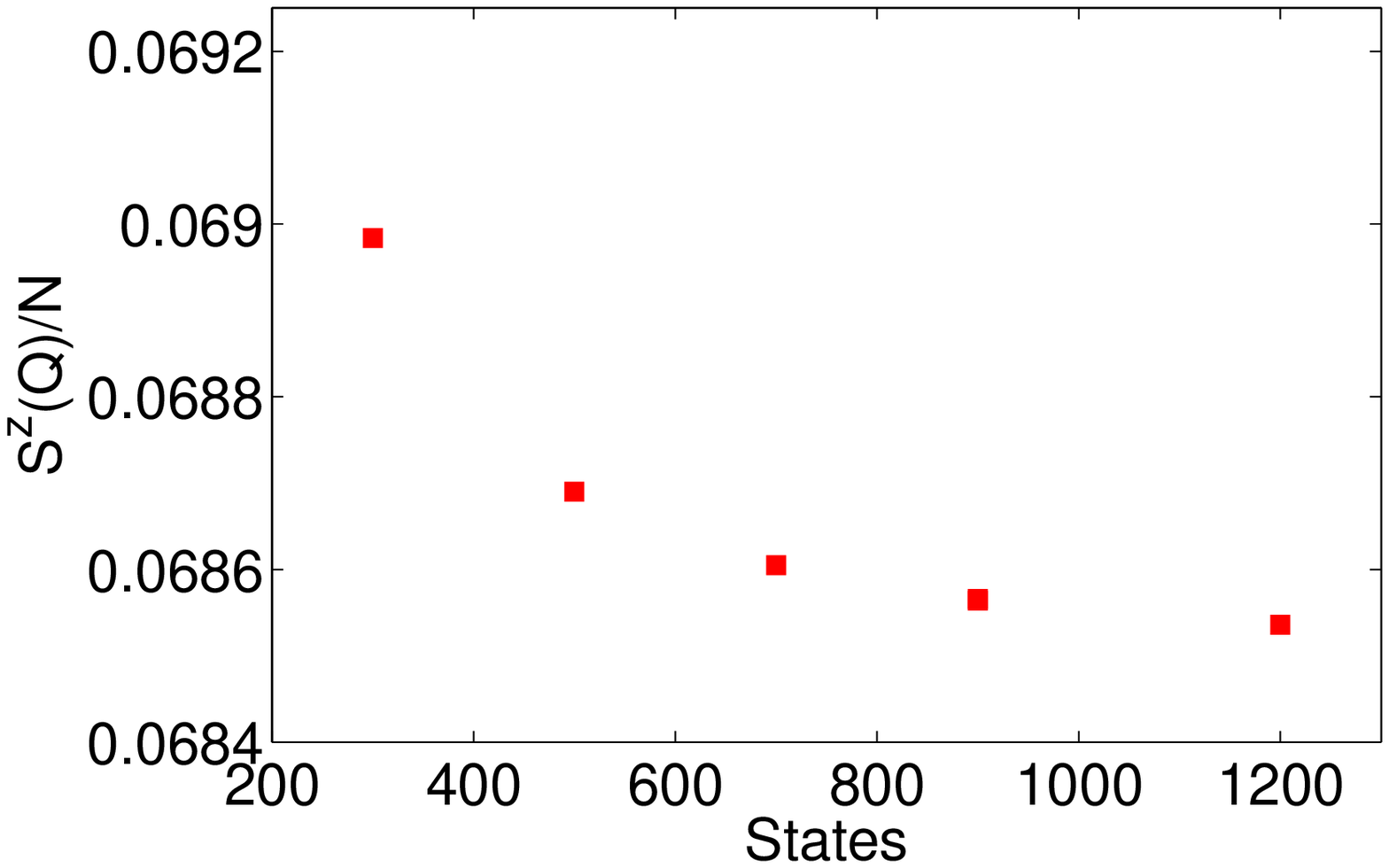}
\includegraphics[width=.8\columnwidth]{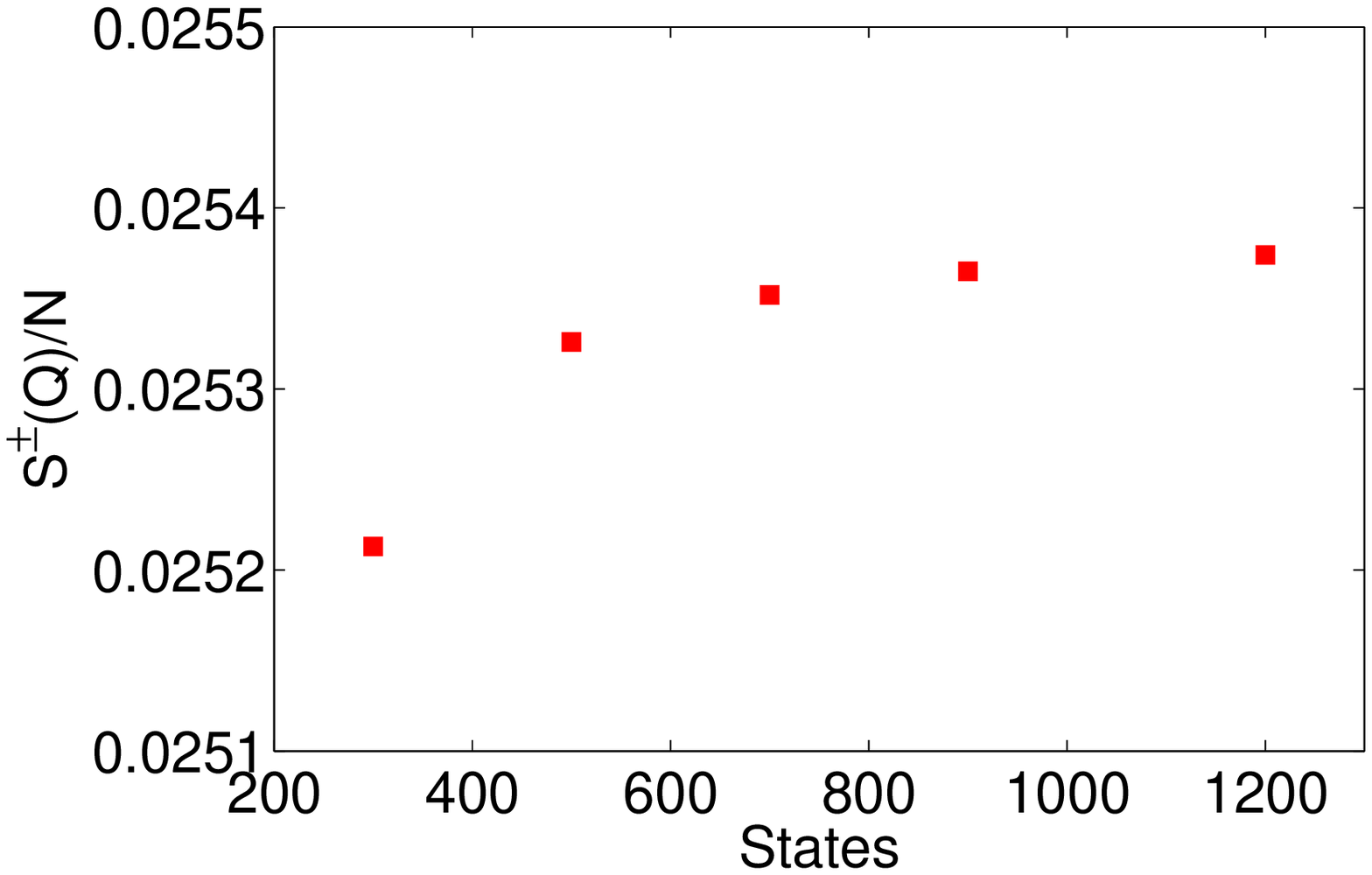}
\caption{The structure factors $S^z(Q)/N$ (upper panel) and $S^{\pm}(Q)/N$ (lower panel) 
as a function of the number of kept states $m$ for $J/J_z=1$, $M=1/6$ and $N=6\times6$.
\label{L36sf}}\end{center}
\end{figure}

The following Figs.~\ref{L36e}-\ref{L108} illustrate the typical behavior of
the DMRG data for the energy, the structure factors, and quantum information measures at
examplary values of the parameters and different system sizes as a function 
of number of states kept.  Note, that the energies alone determine much of the 
phase diagram since they define the magnetization plateau of the 1/3 N\'eel phase.

\begin{figure}[t]
\begin{center}
\includegraphics[width=.8\columnwidth]{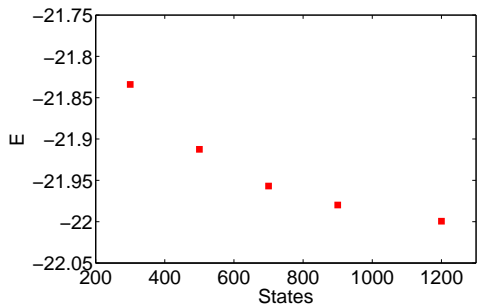}
\caption{The energy $E$ dependence of the kept states at $J/J_z=1$, $M=1/6$ and $N=6\times9$.
\label{L54e}}\end{center}
\end{figure}

\begin{figure}[t]
\begin{center}
\includegraphics[width=0.8\columnwidth]{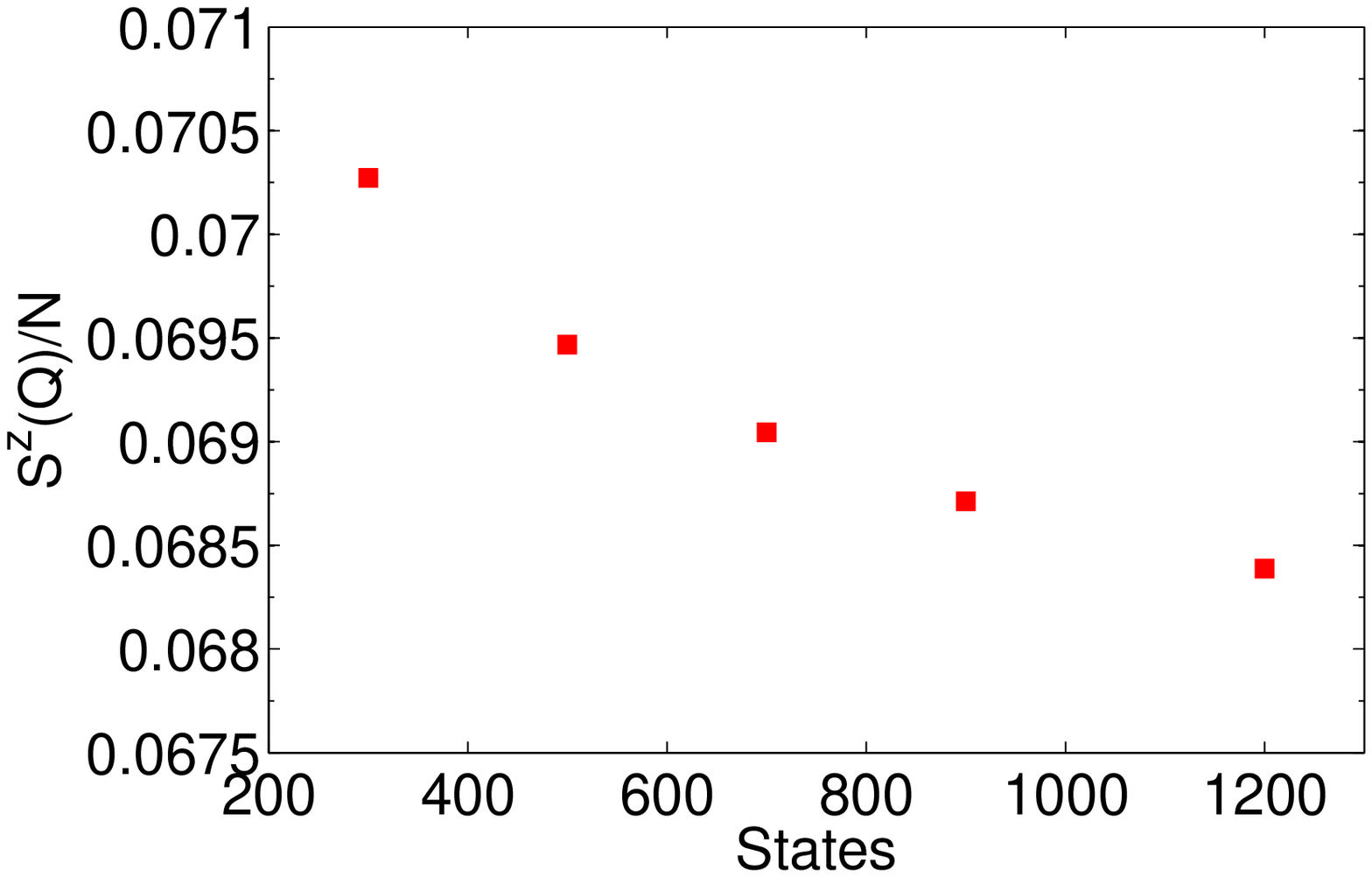}\\
\includegraphics[width=0.8\columnwidth]{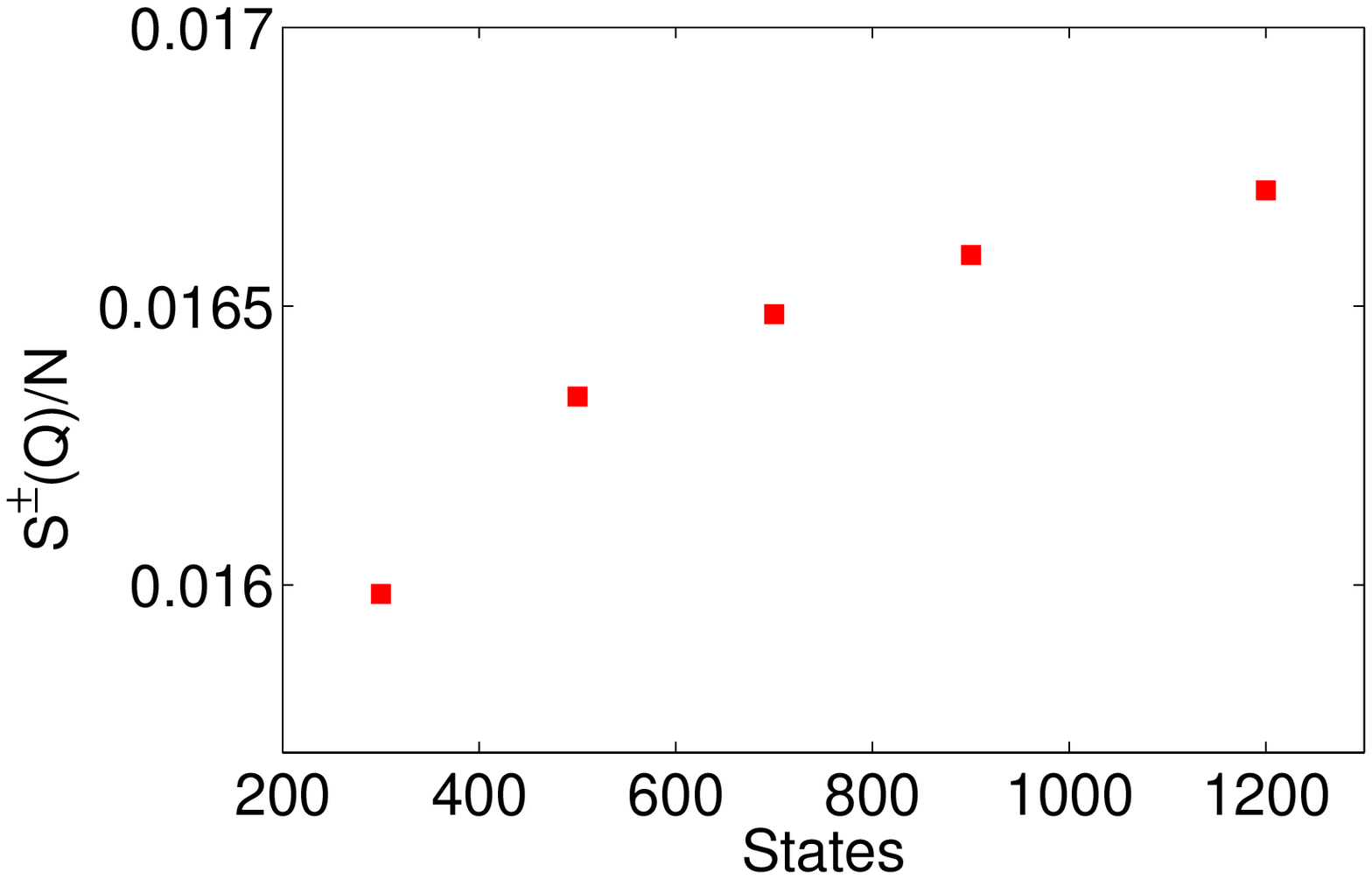}
\caption{The structure factors $S^z(Q)/N$ (left panel) and $S^{\pm}(Q)/N$ (right panel) dependence of the kept states at $J/J_z=1$, $M=1/6$ and $N=6\times9$.
\label{L54sf}}\end{center}
\end{figure}
The behavior of the energy and structure factor is shown in Fig.~\ref{L36e}-\ref{L54sf}
for system sizes 6x6 and 6x9. As expected 
the accuracy improves with number of states kept and is fully sufficient already 
for ca.~1200 kept states. 
In particular, the
difference between the data for 800 or 1200 kept states would not be noticable in
any of the plots. More importantly, there are no big jumps which would be
an indicator for metastable states.

Larger system sizes of $9\times 12$ are more difficult. As shown in Fig.~\ref{L108} there may be 
large jumps when going from 600 states to 900 states, which indicates a metastable 
situtation.  However, all paramaters converge to  stable values for larger values
of the number of states kept.  We have checked convergence for up to $m=2400-3000$ 
at selected points.

\begin{figure}[h]
\begin{center}
\includegraphics[width=0.8\columnwidth]{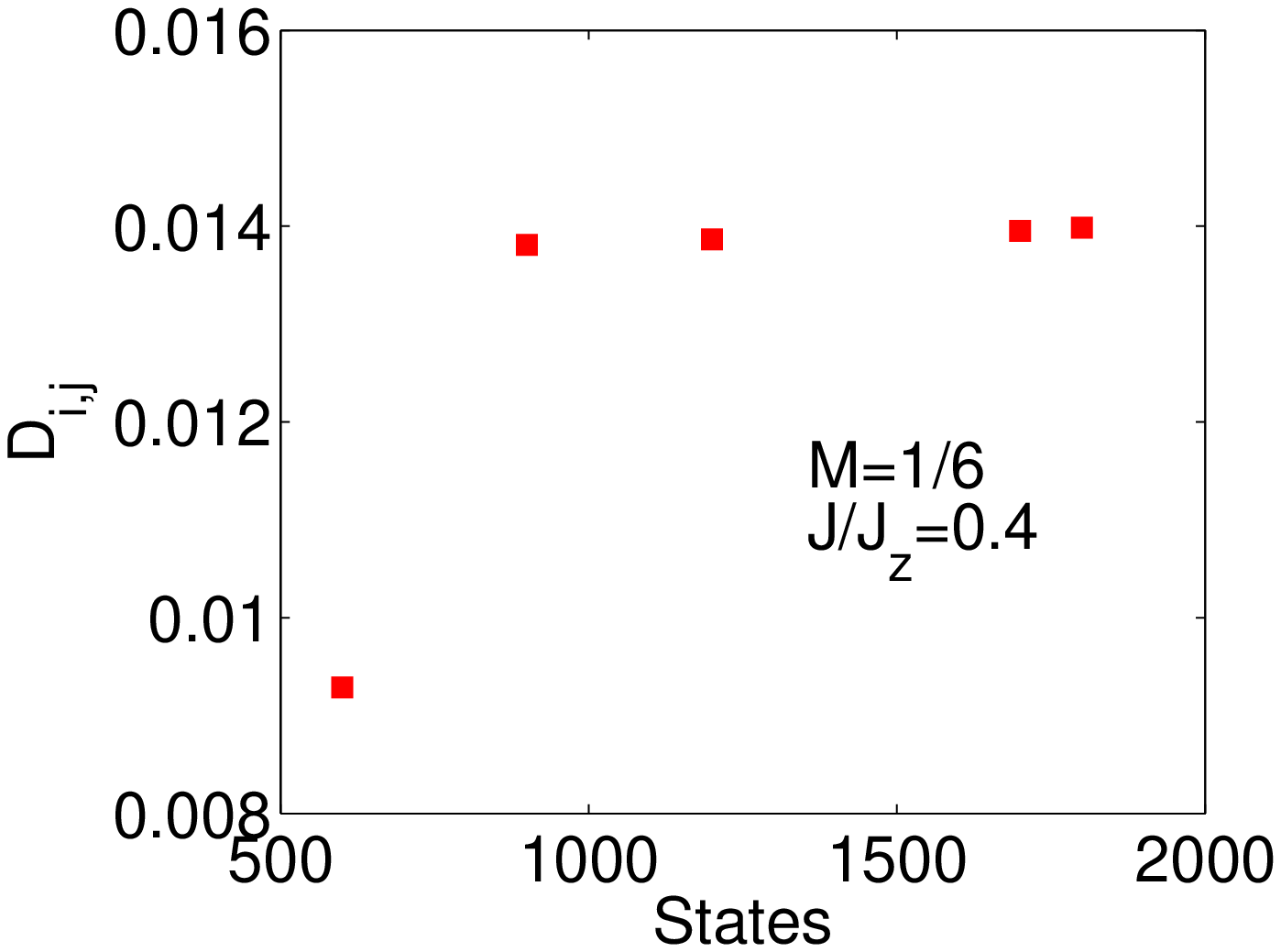}
\includegraphics[width=0.8\columnwidth]{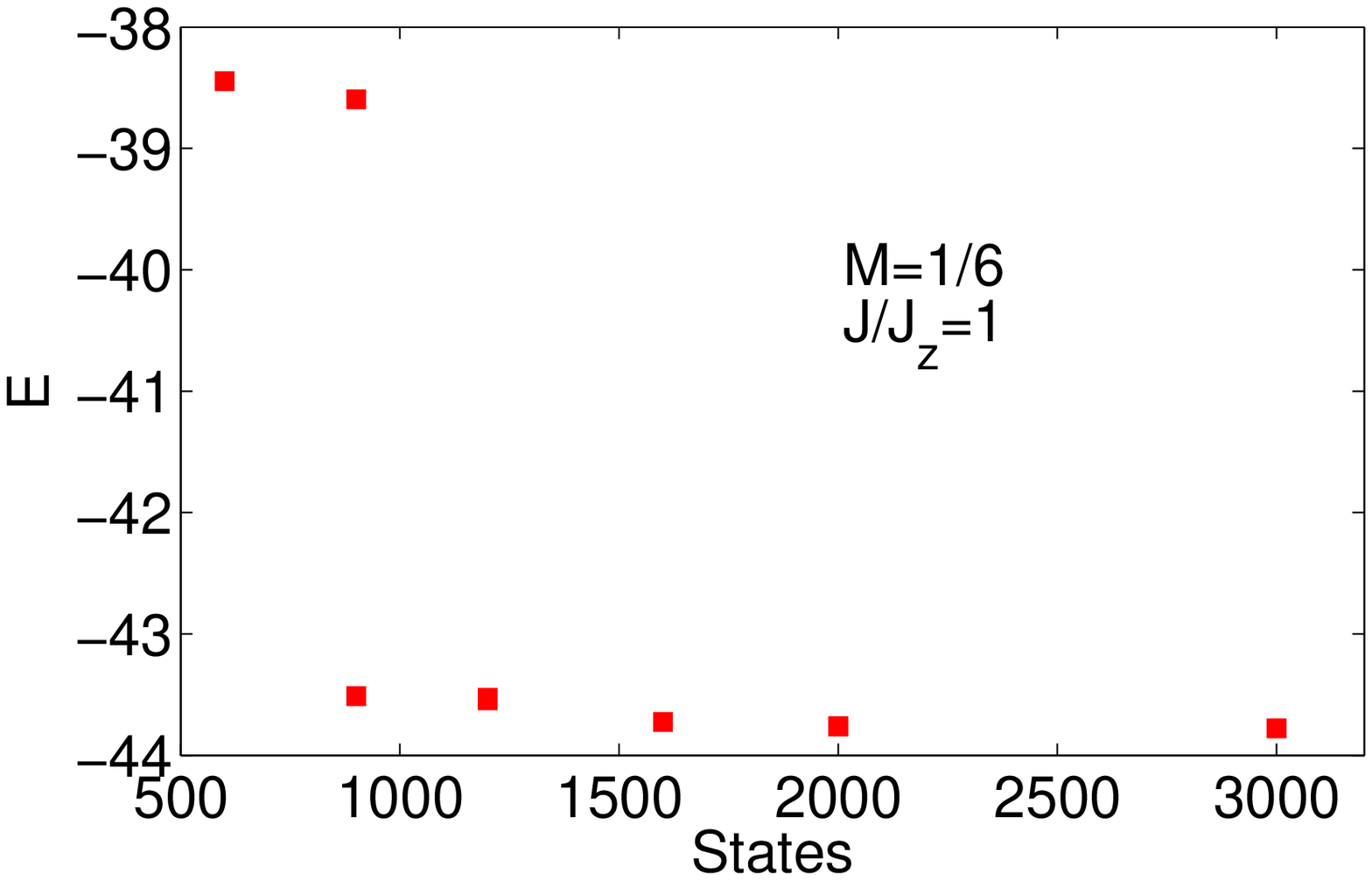}
\includegraphics[width=0.8\columnwidth]{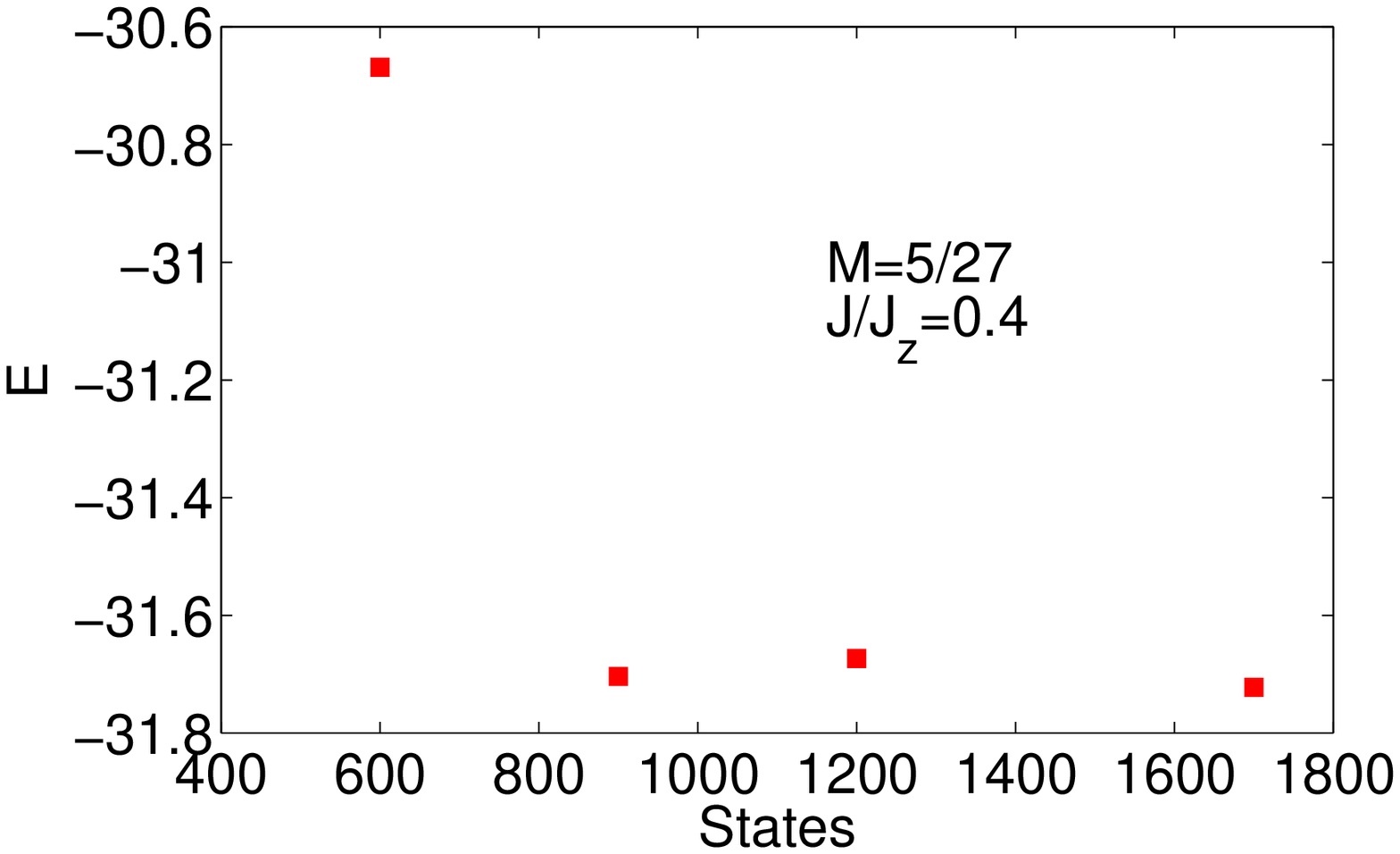}
\caption{Energy and quantum discord as a function of number of states kept in a $9\times 12$ lattice for different phases in the phase diagram and different parameters as indicated in the legends.
\label{L108}}\end{center}
\end{figure}
\bibliographystyle{apsrev}

\bibliographystyle{apsrev}

\begin{thebibliography}{100}
\bibitem{balents10}
L. Balents, Nature (London) {\bf 464}, 199 (2010).


\bibitem{anderson}
P.W.~Anderson, Mater. Res. Bull. {\bf 8}, 153 (1973).


\bibitem{sorella99} L. Capriotti, A. E. Trumper, and S. Sorella,
Phys. Rev. Lett. {\bf 82}, 3899 (1999).


\bibitem{ed} T. Sakai and H. Nakano, Phys. Rev. B {\bf 83}, 100405(R) (2011).


\bibitem{ccm}
D. J. J. Farnell, R. Zinke, J. Schulenburg, and J. Richter, J.
Phys. Condens. Matter {\bf 21}, 406002 (2009).


\bibitem{tri11}
T. Susuki, N. Kurita, T. Tanaka, H. Nojiri,
A. Matsuo, K. Kindo, and H. Tanaka
\newblock Phys.~Rev.~Lett. {\bf 110}, 267201 (2013);


\bibitem{tri10}
Y. Shirata, H. Tanaka, A. Matsuo, and K. Kindo
\newblock Phys.~Rev.~Lett. {\bf 108}, 057205 (2012)


\bibitem{tri12}
H.D. Zhou, C. Xu, A.M. Hallas, H.J. Silverstein, C.R. Wiebe, I. Umegaki, J.Q. Yan,
T.P.  Murphy, J.-H. Park, Y. Qiu, J.R.D. Copley, J. S. Gardner, and Y. Takano,
Phys. Rev. Lett. {\bf 109}, 267206 (2012).


\bibitem{tri13}
G. Koutroulakis, T. Zhou, C.D. Batista, Y. Kamiya, J.D. Thompson, S.E. Brown, and
H.D. Zhou, preprint arXiv:1308.6331 unpublished (2013).


\bibitem{watarai01} S. Watarai, S. Miyashita and H. Shiba,
J. Phys. Soc. Jpn. {\bf 70}, 532 (2001).


\bibitem{tocchio13}
L.~F.~Tocchio, H.~Feldner, F.~Becca, R.~Valent\'i, and C.~Gros,
Phys. Rev. B {\bf 87}, 035143 (2013).


\bibitem{balents13}
R. Chen, H. Ju, H.-C. Jiang, O. A. Starykh, and L. Balents, Phys. Rev. B {\bf 87}, 165123 (201
3).


\bibitem{white11}
A. Weichselbaum and S.R. White, Phys. Rev. B {\bf 84}, 245130 (2011);
K. Harada, Phys. Rev. B {\bf 86}, 184421 (2012).

\bibitem{tocchio14} L.F. Tocchio, C. Gros, X-F. Zhang, S. Eggert,
\prl {\bf 113}, 246405 (2014).


\bibitem{tri} S. Wessel and M. Troyer, Phys. Rev. Lett. {\bf 95}, 127205
(2005); D. Heidarian and K. Damle, Phys. Rev. Lett. {\bf 95},
127206(2005); R. G. Melko, A. Paramekanti, A. A. Burkov,
A.Vishwanath, D. N. Sheng, and L. Balents, Phys. Rev.Lett. {\bf
95}, 127207 (2005); M. Boninsegni and N. Prokof'ev ,
Phys.~Rev.~Lett. {\bf 95}, 237204 (2005).


\bibitem{moessner08} A.~Sen, P.~Dutt,
K.~Damle, and R.~Moessner,\newblock Phys. Rev. Lett. {\bf 100},
147204 (2008).


\bibitem{imp}
X. F. Zhang, Y.C. Wen, and S. Eggert, Phys.~Rev.~B {\bf 82},
220501(R) (2010).


\bibitem{tri_af}
F.~Wang, F.~Pollmann, and A.~Vishwanath,
\newblock Phys.~Rev.~Lett. {\bf 102}, 017203 (2009).
D.~Heidarian and A.~Paramekanti,
\newblock Phys.~Rev.~Lett. {\bf 104}, 015301 (2010).

\bibitem{tri_1st}
D. Yamamoto, I. Danshita, and C.A.R. S\'{a} de Melo
Phys.~Rev.~A {\bf 85}, 021601(R) (2012); L. Bonnes and S.
Wessel, Phys.~Rev.~B {\bf 84}, 054510 (2011); D. Yamamoto,
T. Ozaki, C.A.R. S\'{a} de Melo, I. Danshita
Phys. Rev. A {\bf 88}, 033624 (2013).


\bibitem{tri_sc}
X.-F. Zhang, R. Dillenschneider, Y. Yu, and S. Eggert
\newblock Phys. Rev. B {\bf 84}, 174515 (2011);


\bibitem{miyashita10} S. Miyashita, 
Proc. Jpn. Acad., Ser. B {\bf 86}, 643
(2010).


\bibitem{seabra11}
L. Seabra, T. Momoi, P. Sindzingre, and N. Shannon, Phys. Rev. B {\bf 84}, 214418 (2011).


\bibitem{kawamura85}
H. Kawamura and S. Miyashita, J. Phys. Soc. Jpn. {\bf 54}, 4530 (1985).

\bibitem{classical} S. Miyashita, J. Phys. Soc. Jpn. {\bf 55}, 3605 (1986).


\bibitem{xxz}
D.~Yamamoto, G.~Marmorini, I.~Danshita, {Phys. Rev. Lett. {\bf 112} 127203 (2014).}


\bibitem{yamamoto2} 
D.~Yamamoto, G.~Marmorini, I.~Danshita, Phys. Rev. Lett. {\bf 112}, 259901(E) (2014).


\bibitem{starykh14}  O.A. Starykh, W. Jin, A.V. Chubukov, \prl {\bf 113}, 087204 (2014).





\bibitem{metav14} A. Metavitsiadis, R. Dillenschneider, and S. Eggert,
 Phys. Rev. B {\bf 89}, 155406 (2014). 


\bibitem{yoshikawa04} S. Yoshikawa, K. Okunishi, M. Senda and S. Miyashita,
J. Phys. Soc. Jpn. {\bf 73}, 1798 (2004).


\bibitem{zhang13} X-F. Zhang and S. Eggert
Phys. Rev. Lett. {\bf 111}, 147201 (2013).


\bibitem{tri_afn}
H.C. Jiang, M.Q. Weng, Z.Y. Weng, D.N. Sheng, and L. Balents,
\newblock Phys. Rev. B {\bf 79}, 020409(R) (2009).


\bibitem{jiang08} H.C. Jiang, W.Y. Weng, and D.N. Sheng, \prl {\bf 101}, 117203 (2008).


\bibitem{white}
S.~R. White.
\newblock Phys.~Rev.~Lett. {\bf 69}, 2863, (1992);
E.M. Stoudenmire and S.R. White
\newblock Annu. Rev. Condens. Matter Phys. {\bf 3}, 111 (2012)


\bibitem{schollwoeck}
U.~Schollw\"ock. \newblock Rev.~Mod.~Phys., {\bf 77}, 259 (2005);
\newblock Annals of Physics, {\bf 326}, 96 (2011).


\bibitem{xiang}
T. Xiang, J. Lou, and Z. Su.
\newblock Phys.~Rev.~B, {\bf 64}, 104414 (2001).


\bibitem{yamamoto3} Y. Kato, D. Yamamoto, and I. Danshita, \prl {\bf 112}, 055301 (2014).

\bibitem{highfield}
{For large magnetization exact diagonalization gives
more accurate results than DMRG since the Hilbert space is relatively small.}


\bibitem{sq}
G. G. Batrouni and R. T. Scalettar,
\newblock Phys.~Rev.~Lett. {\bf 84}, 1599 (2000)


\bibitem{anyon}
T. Keilmann, S. Lanzmich, I. McCulloch and M. Roncaglia, Nature
Communication, {\bf 2}, 361 (2011)


\bibitem{gap}
{For system size
$9\times12$ the even magnetization has a better
accuracy, so we choose $B(M) = [E(M+2/N)-E(M)]/2$.}


\bibitem{miyashita86} H. Nishimori, and S. Miyashita,
J. Phys. Soc. Jpn. {\bf 55}, 4448 (1986).


\bibitem{pierre1994}
B. Bernu, P. Lecheminant, C. Lhuillier, and L. Pierre,
Phys. Rev. B {\bf 50}, 10048 (1994).


\bibitem{honecker1999} A. Honecker, J. Phys.: Condens. Matter {\bf 11}, 4967 (1999).


\bibitem{honecker2004} A. Honecker, J. Schulenburg, and J. Richter,
J. Phys.: Condens. Matter {\bf 16}, S749 (2004).


\bibitem{zheng-06}
W. Zheng, J. O. Fj\ae restad, R. R. P. Singh, R. H. McKenzie, and R. Coldea, Phys. Rev. B {\bf
 74}, 224420 (2006).


\bibitem{chubukov91} A.V. Chubukov and D.I. Golosov,
J. Phys.: Condens. Matter {\bf 3}, 69 (1991).


\bibitem{poilblanc09} M. Raczkowski and D. Poilblanc, \prl {\bf 103}, 027001 (2009).

{\bibitem{ee1}
A. Osterloh, L. Amico1, G. Falci and R. Fazio, Nature {\bf 416}, 608 (2002)

\bibitem{ee2}
\"{O}. Legeza and J. S\'{o}lyom, \prl {\bf 96}, 116401 (2006)

\bibitem{ee3}
L. Amico, R. Fazio, A. Osterloh, and V. Vedral, \rmp {\bf 80}, 517 (2008)}
 

\bibitem{Wootters} W.K. Wootters, Phys. Rev. Lett. \textbf{80}, 2245 (1998).


\bibitem{dillen08}
R. Dillenschneider, \newblock \prb {\bf 78}, 224413
(2008).


\bibitem{QD}
H. Ollivier and W.H. Zurek, \newblock
Phys.~Rev.~Lett. {\bf 88}, 017901 (2001).



\end{thebibliography}


%
\end{document}

\bibitem{sengstock} C. Becker, P. Soltan-Panahi, J. Kronj\"ager, S. D\"orscher, K. Bongs,
 and K. Sengstock, New J. Phys. {\bf  12}, 065025 (2010).

\bibitem{tri_nh}A. Eckardt, P. Hauke, P. Soltan-Panahi, C. Becker, K. Sengstock and M. Lewenstein,
Europhys. Lett. {\bf 89}, 10010 (2010)